\documentclass[conference]{IEEEtran}
\usepackage{cite}
\ifCLASSINFOpdf
   \usepackage[pdftex]{graphicx}
  \else
    \usepackage[dvips]{graphicx}
\fi
\usepackage{array}
\usepackage[tight,footnotesize]{subfigure}
\usepackage[font=footnotesize]{subfig}
\usepackage{url}
\begin{document}

\title{Performance of the LBNL FastCCD\\ for the European XFEL}

\author{Friederike Januschek$^a$, Ivana Kla$\mathrm{\check{c}}$kova$^{a,b}$, Nord Andresen$^c$, Peter Denes$^c$, Steffen Hauf$^a$, \\John Joseph$^c$, Markus Kuster$^a$, Craig Tindall$^c$\\
\footnotesize{$^a$ European XFEL GmbH, Holzkoppel 4, 22869 Schenefeld, Germany}\\
\footnotesize{$^b$ Slovak University of Technology, Faculty of Electrical Engineering and Information Technology, Ilkovi$\mathrm{\check{c}}$ova 3, Bratislava, SK-812 19, Slovak Republic}\\
\footnotesize{$^c$ Lawrence Berkeley National Laboratory, 1 Cyclotron Road, Berkeley, CA 94720, U.S.A.}}
\maketitle

\begin{abstract}
%\boldmath
The European X-ray Free Electron Laser (XFEL.EU) is currently being commissioned in Schenefeld, Germany. From 2017 onwards it will provide spatially coherent X-rays of energies between 0.25\,keV and 25\,keV with a unique timing structure.
One of the detectors foreseen at XFEL.EU for the soft X-ray regime (energies below 6\,keV) is a quasi column-parallel readout FastCCD developed by Lawrence Berkeley National Lab (LBNL) specifically for the XFEL.EU requirements. 
Its sensor has 1920$\times$960 pixels of 30\,$\mu$m $\times$30\,$\mu$m size with a beam hole in the middle of the sensor. The camera can be operated in full frame and frame store mode.
With the FastCCD a frame rate of up to 120~fps can be achieved, but at XFEL.EU the camera settings are optimized for the 10\,Hz XFEL bunch-mode.
The detector has been delivered to XFEL.EU. Results of the performance tests and calibration done using the XFEL.EU detector calibration infrastructure are presented quantifying noise level, gain and energy resolution. \end{abstract}

\IEEEpeerreviewmaketitle

\section{Introduction}
From 2017 onwards the European X-ray Free Electron Laser (XFEL.EU) \cite{Altarelli2006}, which is currently being commissioned in Schenefeld, Germany, will provide spatially coherent X-rays with a unique timing structure: there will be up to 2700 pulses, with a 4.5 MHz repetition rate, 10 times per second at very high photon fluxes up to $10^{13}$ photons per pulse \cite{Tschentscher2011}. For the planned energy range between 0.25 keV and 25 keV several dedicated detectors are foreseen \cite{Kuster2014}. One of the detectors for the soft X-ray regime will be a FastCCD with a quasi column-parallel readout, which has been developed at the Lawrence Berkeley National Lab (LBNL)~\cite{Denes2009}. 

\section{The LBNL FastCCD}
\begin{figure}[h!]
\centering
	%\vspace{-.5cm}
	\includegraphics[width=0.6\columnwidth, angle=90]{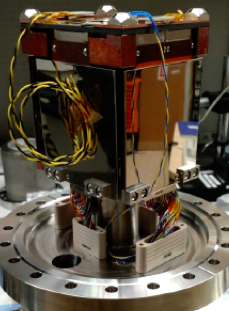}
	\caption{\it The assembled FastCCD camera for XFEL.EU.}
    \label{fig:camera}
\end{figure}

The 1k Frame Store (FS) FastCCD Sensor System developed at LBNL is a soft X-ray detector  
 including a readout system optimized for experiments at advanced light sources.  It is designed for energies of $0.25\;\mathrm{keV}<E_{\gamma}<6\;\mathrm{keV}$. The main components of the system are an in-vacuum camera head shown in Figure~\ref{fig:camera}, a camera interface electronics board and an ATCA based readout and data processing subsystem~\cite{Denes2012}. The full camera head assembly includes the three separate electronics subassemblies and a cooling system to cool the camera to operating temperature of approximately -57$^\circ$C. For increased speed the readout is done almost column parallel with a ten column multiplexed readout. 

The detector manufactured for XFEL.EU has a sensor consisting of 1920$\times$960~pixels of a size of $30\,\mathrm{\mu m} \times30\,\mathrm{\mu m}$. Two operating modes are implemented: one using the entire sensor area for imaging (full-frame mode), the second using only the central 960$\times$960~pixels for imaging and the remainder as a frame-store area. The detector specifications are given in Table~\ref{table:specs}.
The maximum achievable frame rate is 120 frames per second (fps) in frame-store mode and 60~fps in full-frame mode.
At XFEL.EU the camera will mainly be operated at 10~Hz, corresponding to the pulse train rate of the FEL.   

As the intensity of the XFEL.EU beam at the detector may be very high, a central hole for the beam to pass through is required. This hole of the size 1.8~mm was laser-drilled after the sensor had been fabricated, bonded and tested. During the drilling the sensor was protected by a thin layer of First Contact\texttrademark~cleaner.%\cite{FirstCon}.    

\begin{table}[t]
	\centering
	\caption{ \it Technical specifications of the FastCCD detector developed at LBNL for XFEL.EU.}
	%\begin{tabular}{llll}
		\vspace{0.3cm}
	%\begin{tabular}{p{.45\columnwidth}p{.6\columnwidth}p{.35\columnwidth}p{.30\columnwidth}}
	\begin{tabular}{p{.42\columnwidth}p{.58\columnwidth}}
		%\toprule
		\textbf{Properties}  & \textbf{FastCCD}         \\
		%\midrule
		Photon energy range  & $0.25\,\mathrm{keV}-6\,\mathrm{keV}$ \\
		Pixel size           & $30\,\mathrm{\mu m}\times30\,\mathrm{\mu m}$\\
		Sensor size          & $1920\times960$ pixels \\
		Sensor thickness			 & $200\,\mathrm{\mu m}$  \\
		Dynamic range        & $10^3$ above $0.5\,\mathrm{keV}$\\
		Beam hole size & $1.8\,\mathrm{mm}$ \\
		Speed& 60$\,$fps (FF), 120$\,$fps (FS)\\
		Quantum efficiency& $\geq 94\%$  for $E_{\gamma}>1\;\mathrm{keV}$\\ 
		Vacuum operation & 10$^{-7}$-10$^{-6}$ mbar\\
		%\bottomrule
	\end{tabular} 
\vspace{-0.3cm}
	\label{table:specs}

\end{table}

%\pagebreak
\section{Performance of the FastCCD}
First performance tests have been done showing good performance also after the 1.8~mm hole was cut in the sensor. 
\subsection{Dark measurements}
\begin{figure}[t!]
	\vspace{-.5cm}
	\includegraphics[width=1\columnwidth]{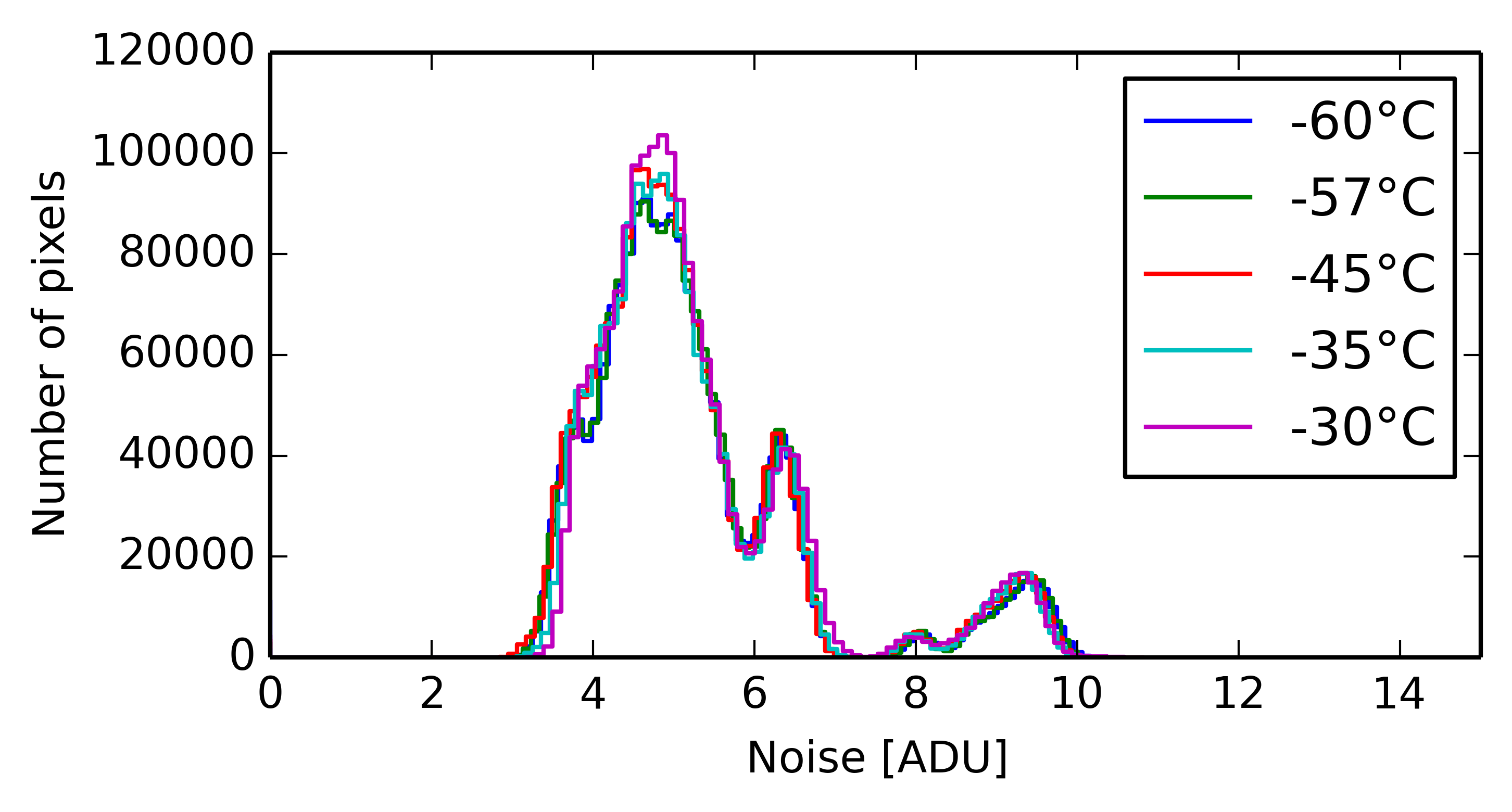}
	\caption{\it Histogram showing the noise for the FastCCD pixels at different temperatures.} %(standard deviation in ADUs for  offset-corrected dark images).}
    \label{fig:noisehisto}
\end{figure}

\begin{figure}[t!]
	\vspace{-.5cm}
	\includegraphics[width=1\columnwidth]{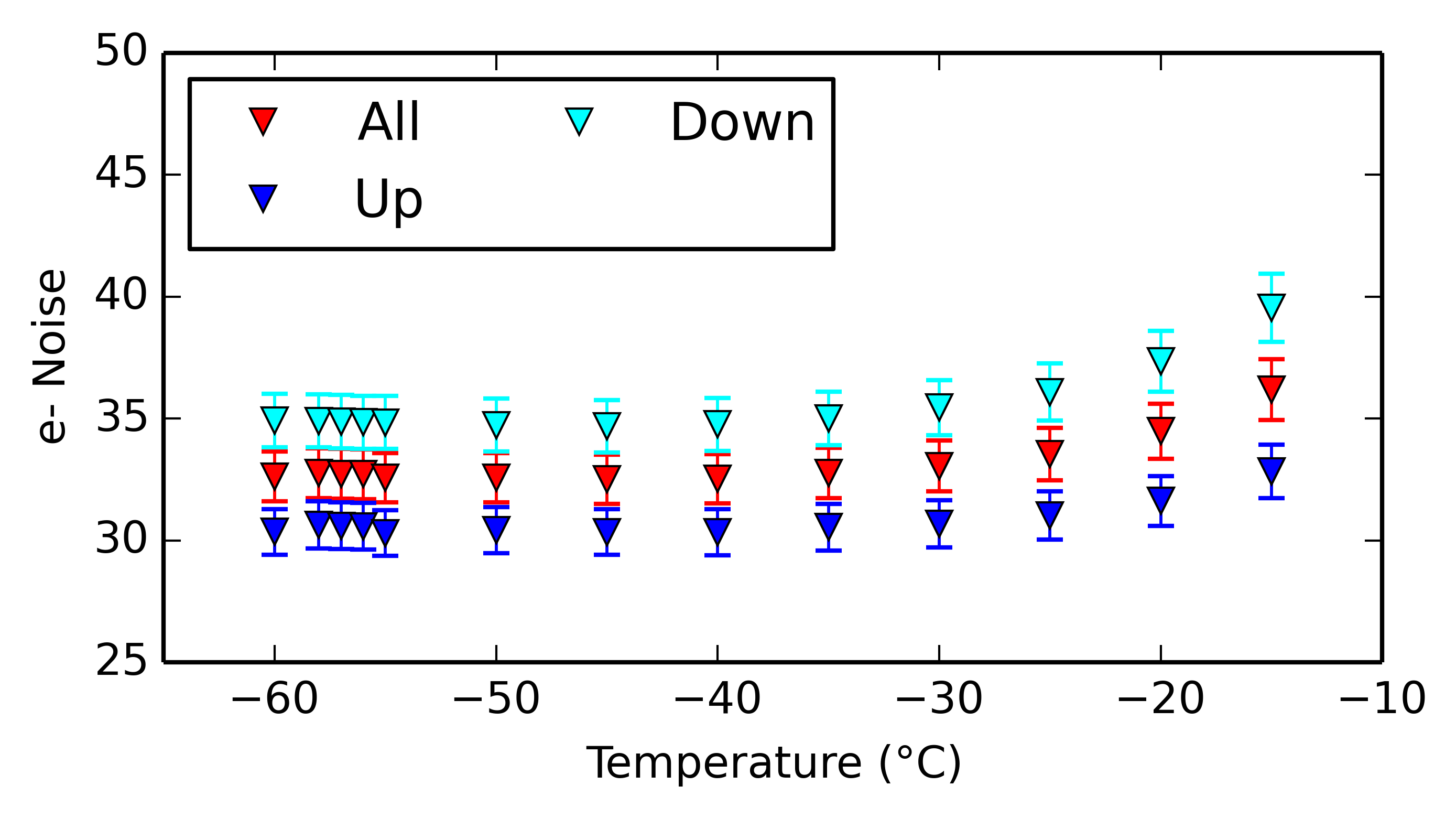}
	\caption{\it Dependance on sensor temperature of the average noise for the whole sensor (red), the upper hemisphere (blue) and the lower hemisphere (cyan).} 
	    \label{fig:noisetemp}
\end{figure}

The RMS noise level was measured to be 5.37$\pm$1.41\,ADU counts and is depicted in the histogram shown in Figure~\ref{fig:noisehisto}. Given the detector settings this is equivalent to $32.7\pm 8.5\,\mathrm{e^-}$. The noise and offset values were found to be stable within about 0.5\% between -60$^{\circ}$C and -30$^{\circ}$C and then start increasing with rising temperature as expected (see Figure~\ref{fig:noisetemp}). 

\subsection{Measurements with Fe-55 illumination}
Initial estimates of the detector response have been done by illuminating the sensor with a 1.1\,GBq Fe-55 source using the XFEL.EU detector calibration laboratory \cite{Sztuk-Dambietz2013}. 

\begin{figure}[t!]
	\vspace{-.5cm}
	\includegraphics[width=0.92\columnwidth]{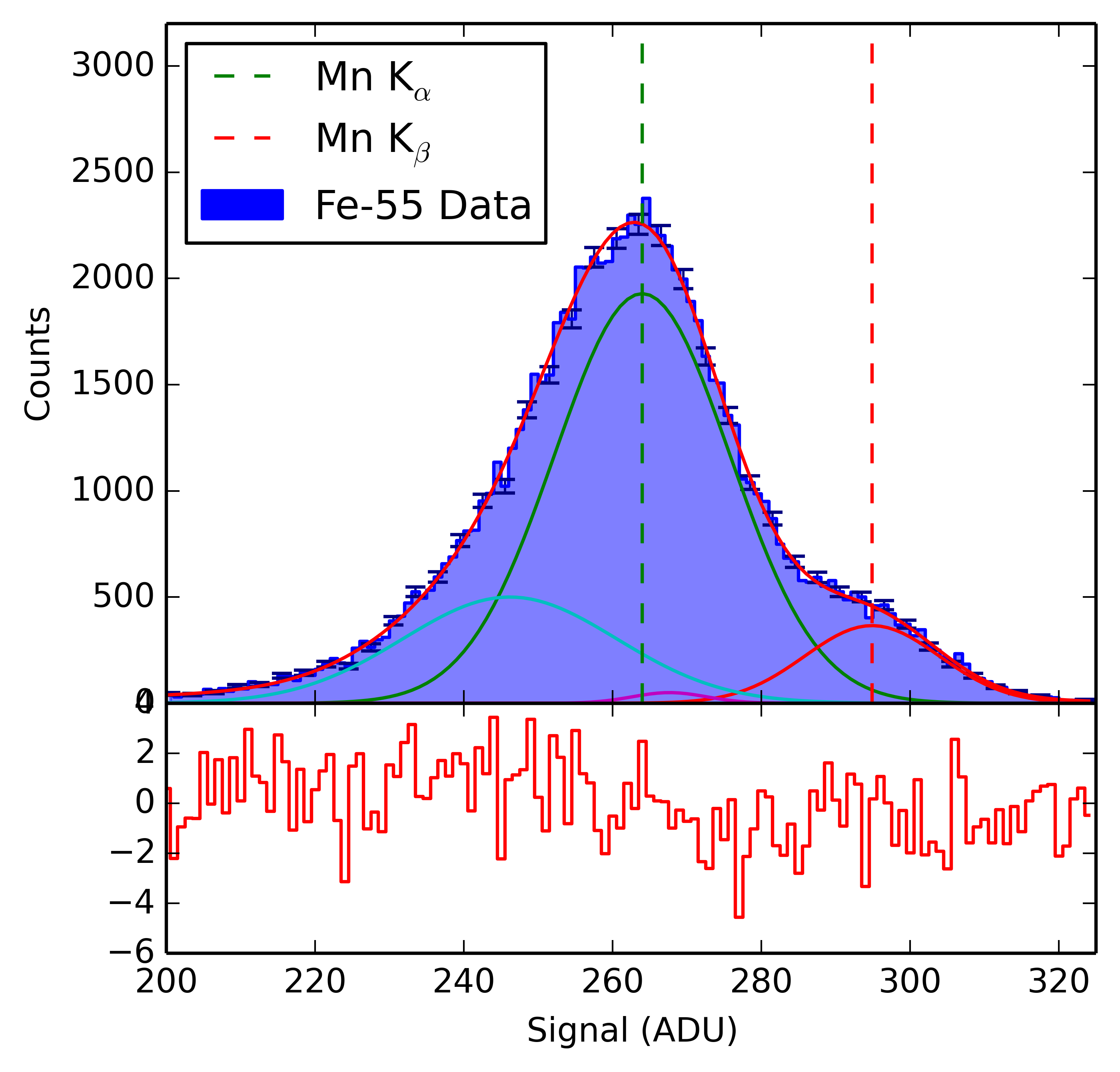}
	\caption{\it The measured energy spectrum of an Fe-55 source is shown, taking only single-pixel events into account. The Mn $K_{\alpha}$ peak is clearly visible. The fit to Mn $K_{\alpha}$ and Mn $K_{\beta}$ and their positions are indicated by lines.}
    \label{fig:singles}
\end{figure}

\begin{figure}[b!]
	\vspace{-.5cm}
	\includegraphics[width=1\columnwidth]{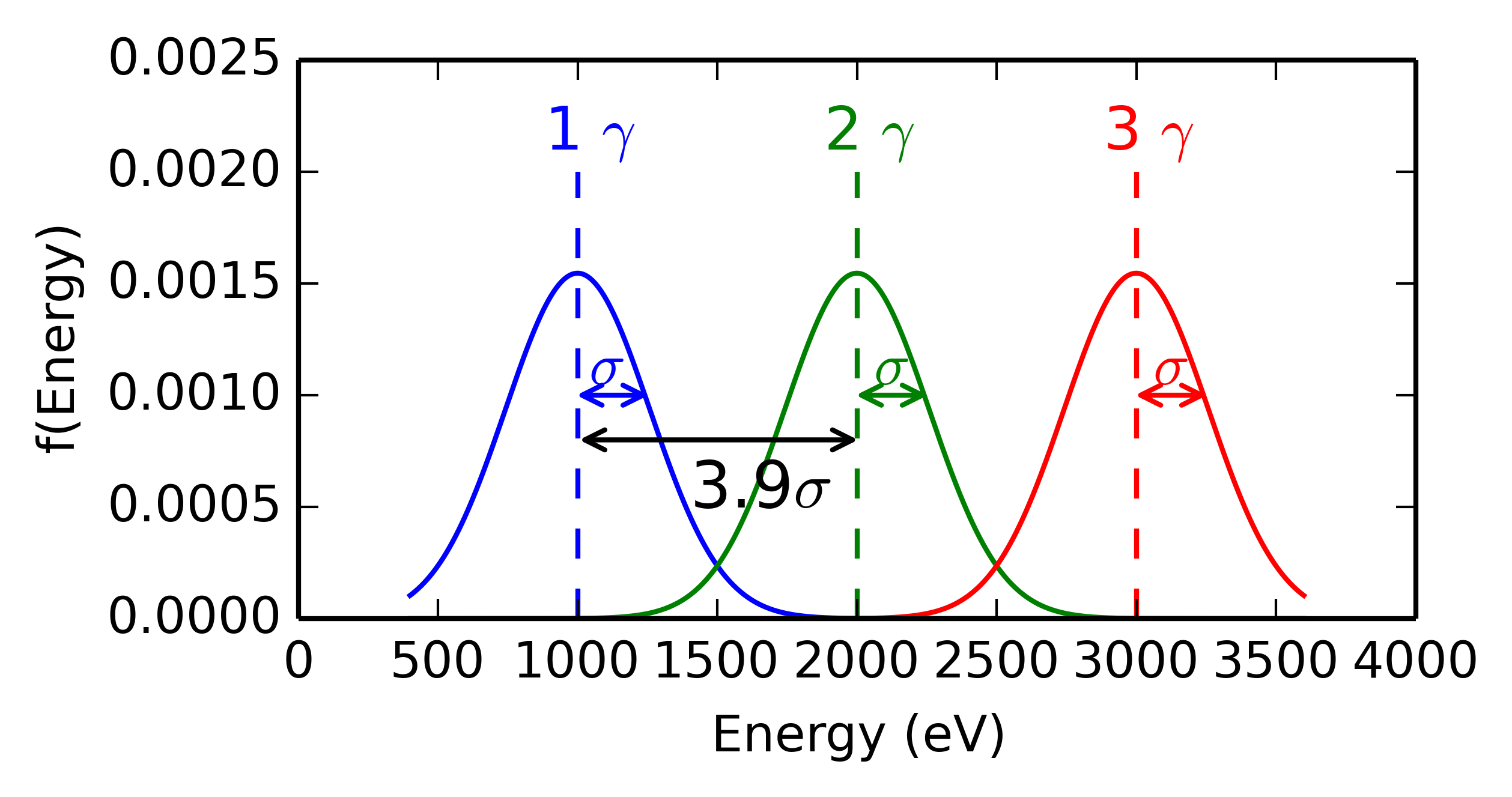}
	\caption{\it Separation power of the FastCCD for photons at 1~keV. The predicted distributions of 1\,$\gamma$, 2\,$\gamma$ and 3\,$\gamma$ events are shown and their separation of 3.9\,$\sigma$, i.e. an overlap of 2.6\% of the event distributions. }
    \label{fig:separation}
\end{figure}

Figure~\ref{fig:singles} shows an energy spectrum of 1-pixel single events (a single pixel having a signal above 5~$\sigma_{noise}$, all neighboring pixels showing no signal, i.e. counts below 3~$\sigma_{noise}$). The Mn~$K_{\alpha}$ is clearly visible and its position as well as the position of the Mn$K_{\beta}$ have been fitted to a 5-gaussian model. The fit to the position of these peaks yields a gain estimate of \mbox{$22.26\pm0.2\;\mathrm{eV/ADU}$}, which is in line with expectations. Note that the data from which the spectrum was produced has not yet been CTI or relative gain corrected.
Nevertheless, an energy resolution of $\sigma=262\pm4\;\mathrm{eV}$ can be observed, corresponding to a peak separation of 3.9\,$\sigma$ at the lower energy range of 1\,keV, which is visualized in Figure~\ref{fig:separation} . The device thus offers reasonable performance for operation in a photon-integrating mode in these energy regimes.

\section{Outlook}
With CTI and relative gain corrections the energy resolution has been improved to up to $\sigma=174\pm17\;\mathrm{eV}$ for part of the detector. These corrections are currently limited by a synchronization issue in the firmware of the FastCCD. With a firmware update imminent, and further optimizations at LBNL (including an exchange of a noisy digitizer board),  performance tests and calibration will continue in the XFEL.EU detector calibration laboratory. Using a pulsed X-ray source is planned, aiming at a calibration that allows an improved energy resolution. 

\clearpage
Afterwards, the FastCCD is to be integrated in the beam line of the Spectroscopy and Coherent Scattering \cite{Scherz2013} (SCS) experiment at XFEL.EU. There it will be utilized for single-shot non-linear imaging experiments and as a secondary diffraction plane for time-resolved holography together with the DSSC detector~\cite{Porro2012}.

\vspace{0.5cm}

\section{Conclusion}
Results of the performance tests for the LBNL FastCCD developed for the European XFEL have been presented. The RMS noise level is $32.7\pm 8.5\,\mathrm{e^-}$. With Fe-55 the energy resolution was measured to be  $\sigma =262\pm 4\,\mathrm{eV}$. This is equivalent to a projected separation power of 3.9\,$\sigma$ at 1\,keV. A further performance improvement is expected after replacement of a noisy digitizer board as well as optimization of the firmware and subsequently a full set of corrections including CTI and relative gain corrections.

\end{document}